\documentclass[aps,pre,twocolumn,superscriptaddress,showpacs,floatfix]{revtex4-1}

\usepackage{graphicx}
\usepackage{dcolumn}
\usepackage{bm}
\usepackage{url}
\usepackage{times}

\usepackage{amsmath}

\usepackage{appendix}
\usepackage[bf]{subfigure}
\usepackage[utf8]{inputenc}

\usepackage{tikz}

\usepackage{xcolor}

\begin{document}


\title{The role of time scale in the spreading of asymmetrically interacting diseases}

\author{Paulo Cesar Ventura}
\affiliation{Instituto de F\'{i}sica de  S\~{a}o Carlos, Universidade de S\~{a}o Paulo, S\~{a}o Carlos, SP, Brazil.}

\author{Yamir Moreno}
\affiliation{Institute for Biocomputation and Physics of Complex Systems (BIFI), University of Zaragoza, 50018 Zaragoza, Spain}
\affiliation{Department of Theoretical Physics, University of Zaragoza, 50018 Zaragoza, Spain}
\affiliation{ISI Foundation, Via Chisola 5, 10126 Torino, Italy}

\author{Francisco A. Rodrigues}
\affiliation{Instituto de Ci\^{e}ncias Matem\'{a}ticas e de Computa\c{c}\~{a}o, Universidade de S\~{a}o Paulo, S\~{a}o Carlos, SP, Brazil.}


\begin{abstract}
Diseases and other contagion phenomena in nature and society can interact asymmetrically, such that one can benefit from the other, which in turn impairs the first, in analogy with predator-prey systems. Here, we consider two models for interacting disease-like dynamics with asymmetric interactions and different associated time scales. Using rate equations for homogeneously mixed populations, we show that the stationary prevalences and phase diagrams of each model behave differently with respect to variations of the relative time scales. We also characterize in details the regime where transient oscillations are observed, a pattern that is inherent to asymmetrical interactions but often ignored in the literature. Our results contribute to a better understanding of disease dynamics in particular, and interacting processes in general, and could provide interesting insights for real-world applications, most notably, the interplay between the  dynamics of fact-checked and fake news.
\end{abstract}

\date{\today}

\maketitle

\section{Introduction}

Spreading processes are ubiquitous in nature and society. A primary example of these phenomena is the propagation of diseases in human and animal populations \cite{Pastor015, Arruda018}. Despite many recent advances in the theoretical, computational and data-driven modeling of diseases, there are still many scientific problems that remain open. Two of such problems are currently of high relevance. On the one hand, we have many limitations when it comes to understand the effects of mobility restrictions and human behavioral changes on the evolution of a disease \cite{Aleta2020}. Secondly, there are many diseases that rarely evolve in isolation, on the contrary, diverse viral strains or different pathogens can either compete for the susceptible population or establish cooperative interactions, both at a population level or inside a host's organism. This paper focuses on studying spreading processes with the aim of shedding some light into the second kind of challenge. 

Early works on interacting diseases~\cite{elveback1964extension,dietz1979epidemiologic,may1983epidemiology,castillo1996competitive,andreasen1997dynamics} have modeled the dynamics of pathogens or strains that interact competitively, showing that their coexistence is not always possible. In the last years, these models have been adapted to networks~\cite{newman2005threshold,karrer2011competing,funk2010interacting,marceau2011modeling,poletto2015characterising,sahneh2014competitive,wang2012dynamics}, including single layer and multiplex networks~\cite{Arruda018, Cozzo018}, as well as metapopulations~\cite{poletto2015characterising}. These works have showed that the network organization plays a fundamental role on the evolution of the dynamical processes, affecting the epidemic threshold and prevalence (e.g.~\cite{wang2014asymmetrically}). Recently, the focus has also been placed on collaborative contagion, in which there is a positive feedback between diseases~\cite{dodds2005generalized,newman2013interacting,chen2013outbreaks,cai2015avalanche,hebert2015complex,cui2017mutually,chen2017fundamental}. In these models, discontinuous phase transitions, in which the prevalence goes from zero to a finite value abruptly, have been observed, as well as simultaneous stability of two epidemic states. It is worth remarking that although the network topology can generate important phenomena, some essential properties can still be observed in homogeneously mixed populations, as demonstrated by Chen and collaborators \cite{chen2017fundamental}.

General models for interacting diseases have also been developed recently. This is the case of \cite{sanz2014dynamics}, where the authors introduced a generic model for two interacting diseases in multiplex networks, comprising the cases of competitive, collaborative and asymmetrical interactions, for both SIS (\emph{Susceptible-Infected-Susceptible}) and SIR (\emph{Susceptible-Infected-Removed}) compartmental models. They calculated the respective epidemic thresholds and studied the prevalence for both the competitive and the collaborative scenarios. However, the asymmetrical case has remained unexplored.

Asymmetrical interactions occur in many complex systems. For example, there are reported cases of HIV viral load suppression by the presence of some other pathogens \cite{watt2003decrease,moss2002suppression,xiang2009viruses}, which suggests an asymmetrical interaction between HIV and many diseases. Asymmetrical interactions have been studied mostly in the case of the interplay between epidemic and awareness~\cite{funk2009spread,wu2012impact,wang2014asymmetrically,wang2016suppressing,velasquez2017interacting,wang2017epidemic,da2019epidemic}, malicious computer worms and spreading countermeasures~\cite{chen2004impact,goldenberg2005distributive,zhu2013mixing,yang2014pulse}, and antigens and immune system agents~\cite{noh2005asymmetrically,ahn2006epidemic,wu2013superinfection}, among other examples. In the case of disease and awareness, the epidemic stimulates information awareness, which in turn tends to reduce the exposure to disease and therefore also in disease prevalence, configuring an asymmetrical interaction between the two processes. It has been shown that, in general, the awareness can effectively reduce the disease prevalence and increase its threshold~\cite{funk2009spread,wu2012impact}. The epidemic, in turn, can sustain an information outbreak even when the latter is bellow its ``independent'' threshold~\cite{wang2014asymmetrically,wang2016suppressing}. Correlations and structural properties of the underlying multiplex network can also influence this propagation process~\cite{funk2009spread,wang2014asymmetrically,wang2016suppressing}.

As mentioned above, some previous works have examined asymmetrically interacting spreading phenomena. However, this has been done for specific cases, and a general description of these processes is, to the best of our knowledge, still missing. Of special interest is the fact that the interacting dynamics can do so at different time-scales, whose effect has not been deeply investigated yet. Changes in the relative time-scale (i.e., the possibility that one process has a different intrinsic clock concerning the other) typically do not change the qualitative behavior of interacting systems, but often lead to relevant quantitative effects. For example, Karrer and Newman \cite{karrer2011competing} showed that, for competing SIR diseases, the slower pathogen may take advantage, and even win the competition due to its inherent time scale. Oliveira and Dickman \cite{de2017advantage} also showed that the slower species can win a competition in a contact process (CP), including scenarios with periodic and stochastic environmental variations. In the context of asymmetrical interactions, Wu and collaborators \cite{wu2013superinfection} and Poletto et al. \cite{Poletto2013} demonstrated that the epidemic threshold is affected by changes in the recovery rate when the reproductive number (the ratio between the spreading and the recovery rates) is kept constant, which is equivalent to a variation of the time scale. Equivalently, it has also been shown that a slower information awareness may have more impact on the epidemic prevalence than a faster one \cite{wang2017epidemic,da2019epidemic}. Therefore, the development of a more general understanding about the influence of the relative time scales of asymmetrically interacting disease-like models is of great interest.

In this paper, we study two SIS-like models for interacting diseases when they are in an asymmetrically interacting regime over homogeneously mixed populations and continuous-time evolution. We rely on the most simple setup that doesn't involve structured populations to extract the intrinsic properties of the models.  In each model, we assign a parameter $\pi$ that allows us to control the relative time scale between the two dynamics while keeping constant the epidemic forces of each disease. We adopt a descriptive analysis of each model, determining whether different aspects are or aren't influenced by the time scale, including the respective phase diagrams, coexistence and oscillatory behavior. We also compare both models and discuss our results in the light of some previous works.
\section{Epidemic models with asymmetrical interaction}
In what follows, regardless of the model considered, we assume that there are two diseases that interact. Moreover, we consider that disease I is the ``prey'' (it is impaired by the other disease) and disease II is the ``predator'' (it is benefited from the presence of the first disease), in an analogy with the asymmetrical interaction of predator-prey systems. A parameter, $\pi$, controls the relative time scale between the two diseases. It is implemented as follows: the rates of all processes promoted by disease I are multiplied by $1 - \pi$, whereas the rates of disease II processes are multiplied by $\pi$. Making $\pi$ to range between $0$ and $1$, we sweep through scenarios in which disease I is faster ($\pi < 0.5$) or slower ($\pi > 0.5$) than disease II, as well as the balanced case ($\pi = 0.5$). Although $\pi$ does not add a new degree of freedom of the model, the advantages of using this approach are: (i) it allows us to control the relative time scale with a single parameter; (ii) the ``overall rate'' of the system, which can be regarded as $\sim(1 - \pi) + \pi$, is kept constant when varying $\pi$; and (iii) plotting variables as a function of $\pi$ is simple because it is limited between $0$ and $1$. Let us now describe each of the models scrutinized in the rest of the paper.
\subsection{Model A: interacting diseases through susceptibility change}
In this variant, we consider that the presence of one disease impairs the spreading of the second one by changing the individual's susceptibility to catch the other disease. It has been used to describe the dynamics of competing pathogens with partial cross immunity \cite{andreasen1997dynamics} and for collaborative contagion \cite{chen2017fundamental,sanz2014dynamics}, but its asymmetrical version is still largely unexplored. The latter regime is a prototypical model to describe processes in which an epidemic coexists with an information dynamics in which awareness regarding the disease plays a role in the spreading of it. Indeed, our Model A is mathematically similar to models used for that purpose \cite{da2019epidemic}.

\begin{figure}[!tb]
    \begin{center}
    \includegraphics[width=0.9\linewidth]{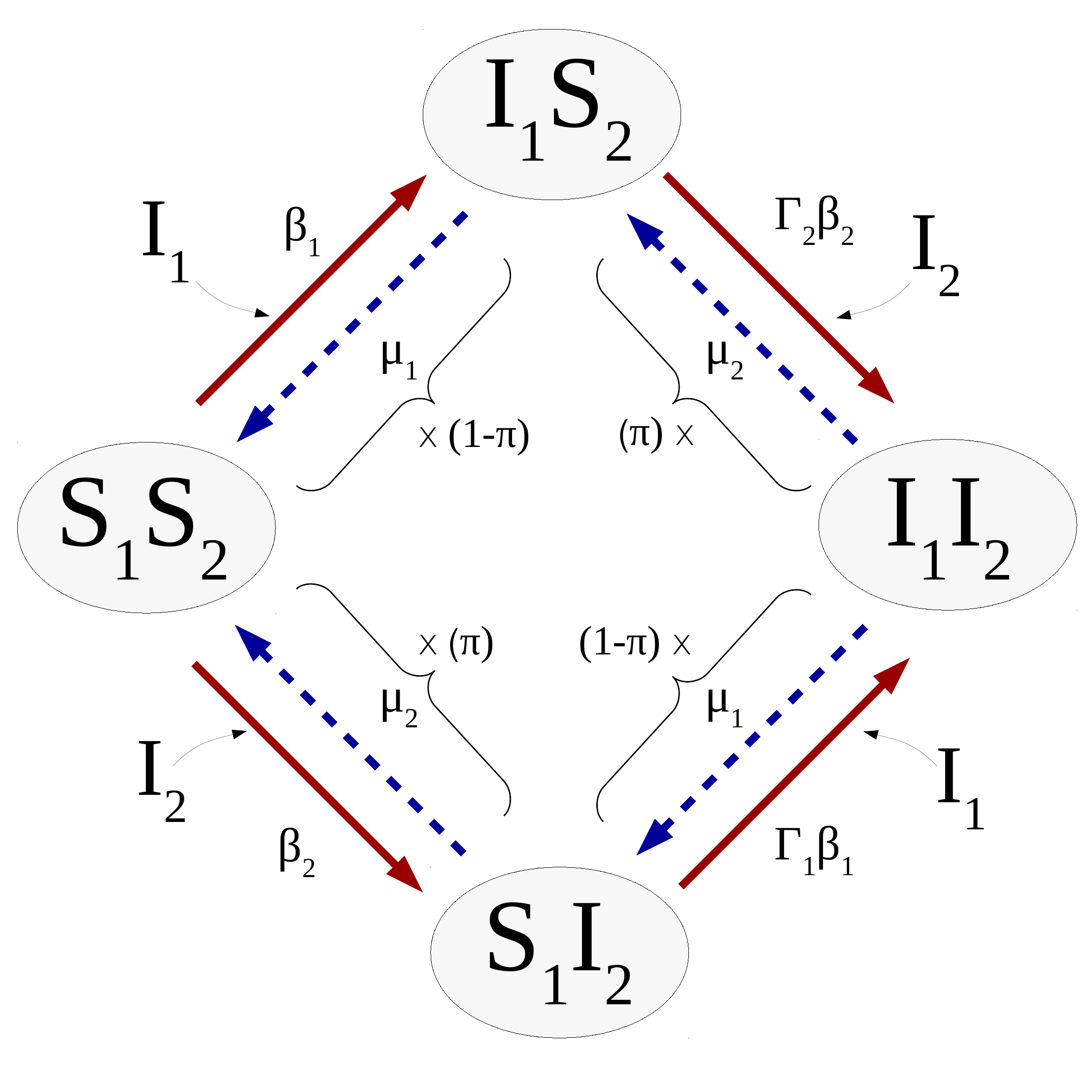}
    \caption{State transitions allowed in model A. The baseline infection and healing rates of disease I (II) are respectively $\beta_1$ ($\beta_2$) and $\mu_1$ ($\mu_2$). $\Gamma_1$ ($\Gamma_2$) represents the modification to the baseline transmission rate of of disease I (II) due to the presence of the other disease in the host. Besides, each rate is multiplied by its corresponding time scale factor: $1 - \pi$ for processes of disease I and $\pi$ for disease II.}
    \label{fig:2sis_scheme}
    \end{center}
\end{figure}

In this model, each individual can either be susceptible to both diseases (S$_1$S$_2$), infected by one disease and susceptible to the second one (I$_1$S$_2$) or (S$_1$I$_2$), or infected by both diseases (I$_1$I$_2$). Note that we denote as I$_1$ and I$_2$ individuals that are, respectively, infected by disease I and II,  regardless of their state with respect to the other disease. For completely susceptible individuals S$_1$S$_2$, the baseline contagion rate of disease I (II) when in contact with an individual infected by disease I (II) is $\beta_1$ ($\beta_2$). For individuals already infected by disease I but susceptible to disease II (I$_1$S$_2$), the contagion rate for disease II is $\Gamma_2\cdot\beta_2$, i.e., it is multiplied by a factor $\Gamma_2$. The same holds for S$_1$I$_2$ individuals, for which the contagion rate by disease I is changed to $\Gamma_1 \cdot \beta_1$. The healing rates from disease I and II are respectively $\mu_1$ and $\mu_2$, and are not affected by the other disease. Figure \ref{fig:2sis_scheme} represents all the possible transitions for this model, with their respective time scale factors as explained before. 

The asymmetrical interaction between the two diseases can be obtained by setting $0 \leq \Gamma_1 < 1$ and $\Gamma_2 > 1$. This means that individuals that hold disease II are less susceptible to catch disease I in comparison to fully susceptible individuals. On the other hand, individuals infected by disease I are more likely to catch disease II. Therefore, disease I enhances the propagation of disease II, whereas disease II impairs the propagation of disease I. We represent the density of individuals in a given state X by $\rho_{x}$, with X being either a composite state (like I$_1$S$_2$) or a simple state (like I$_2$). Based on the diagram from figure \ref{fig:2sis_scheme}, the time evolution of the composite state densities is given by the following equations:

\begin{eqnarray}
    \frac{d\rho_{s_1s_2}}{dt} &=&-(1-\pi) \beta_1 \rho_{s_1s_2}\rho_{i_1} - \pi \beta_2\rho_{s_1s_2}\rho_{i_2} + \nonumber\\ && + (1-\pi)\mu_1\rho_{i_1s_2} + \pi\mu_2\rho_{s_1i_2} \nonumber\\
     \frac{d\rho_{s_1i_2}}{dt} &=&-(1-\pi) \Gamma_1\beta_1 \rho_{s_1i_2}\rho_{i_1} + \pi \beta_2\rho_{s_1s_2}\rho_{i_2} + \nonumber\\ && + (1-\pi)\mu_1\rho_{i_1i_2} - \pi\mu_2\rho_{s_1i_2} \nonumber\\
    \frac{d\rho_{i_1s_2}}{dt} &=& +(1-\pi) \beta_1 \rho_{s_1s_2}\rho_{i_1} - \pi \Gamma_2\beta_2\rho_{i_1s_2}\rho_{i_2} - \nonumber\\&& - (1-\pi)\mu_1\rho_{i_1s_2} + \pi\mu_2\rho_{i_1i_2} \nonumber\\
     \frac{d\rho_{i_1i_2}}{dt} &=& +(1-\pi) \Gamma_1\beta_1 \rho_{s_1i_2}\rho_{i_1} + \pi \Gamma_2\beta_2\rho_{i_1s_2}\rho_{i_2} - \nonumber\\ && - (1-\pi)\mu_1\rho_{i_1i_2} - \pi\mu_2\rho_{i_1i_2}. \nonumber
\end{eqnarray}

The densities are yet subject to the normalization constraint $\rho_{s_1s_2} + \rho_{s_1i_2} + \rho_{i_1s_2} + \rho_{i_1i_2} = 1$. This constraint makes the system effectively three-dimensional. One can reduce the number of equations and simplify the notation by using the following variable change (as done in \cite{chen2017fundamental}):

\begin{equation}
    \begin{cases}
        u = \rho_{i_1} = \rho_{i_1s_2} + \rho_{i_1i_2}, \nonumber\\
        v = \rho_{i_2} = \rho_{s_1i_2} + \rho_{i_1i_2}, \nonumber\\
        w = \rho_{i_1i_2}.\nonumber
    \end{cases}
\end{equation}

For which the dynamical equations are:

\begin{eqnarray}
    \label{eq:2sis_mf_u}
    \Dot{u} &=& (1-\pi) \; [ \beta_1(1 - u) + (\Gamma_1 - 1)\beta_1 (v - w) - \nonumber \\&& \hspace{1.2cm} - \mu_1 ] \; u \\
    \label{eq:2sis_mf_v}
    \Dot{v} &=& \pi \; [ \beta_2(1 - v) + (\Gamma_2 - 1)\beta_2 (u - w) - \mu_2 ] \; v \\
    \Dot{w} &=& (1- \pi)\; [ \Gamma_1 \beta_1 (v - w)u - \mu_1 w ] \; + \; \nonumber \\&& + \pi \; [\Gamma_2 \beta_2 (u - w) v - \mu_2 w]. \label{eq:2sis_mf_w}
\end{eqnarray}

\subsection{Model B: competing diseases with superinfection}

Model B constitutes a modification of models of competing strains, in which a host cannot have the two diseases at the same time. This could be achieved from model A by setting $\Gamma_1 = \Gamma_2 = 0$. However, we also allow the in-host disease replacement via superinfection: if an individual infected by disease I contacts another one infected by disease II, the first can also become infected by disease II, which immediately replaces disease I in the host. The other way around, from disease II to I, is not possible. Superinfection is a phenomenon that is claimed to occur for some diseases such as HIV \cite{fultz1987superinfection,smith2005hiv,ramos2002intersubtype} and bacterial pathogens \cite{feldmeier2002bacterial}, although it does not necessarily leads to in-host replacement of the first infection. There is a considerable amount of works about epidemic models with superinfection, both with homogeneous populations \cite{nowak1994superinfection,iannelli2005strain,martcheva2007vaccine} and complex networks \cite{noh2005asymmetrically,ahn2006epidemic,wu2013superinfection} and including other realistic aspects such as demography. Beyond epidemiological examples, the present model could abstract the interaction between computer viruses and the spreading of anti-malware, or the dynamics of fake news and fact-checking messages. In such scenarios, either the anti-malware or the fact-checked message replaces the previous "infectious agent". We also note that ours is a particular case of the in \cite{wu2013superinfection}, and can also be interpreted as a generalized predator-prey model \cite{brauer2001mathematical}.

\begin{figure}[!tb]
    \begin{center}
    \includegraphics[width=0.8\linewidth]{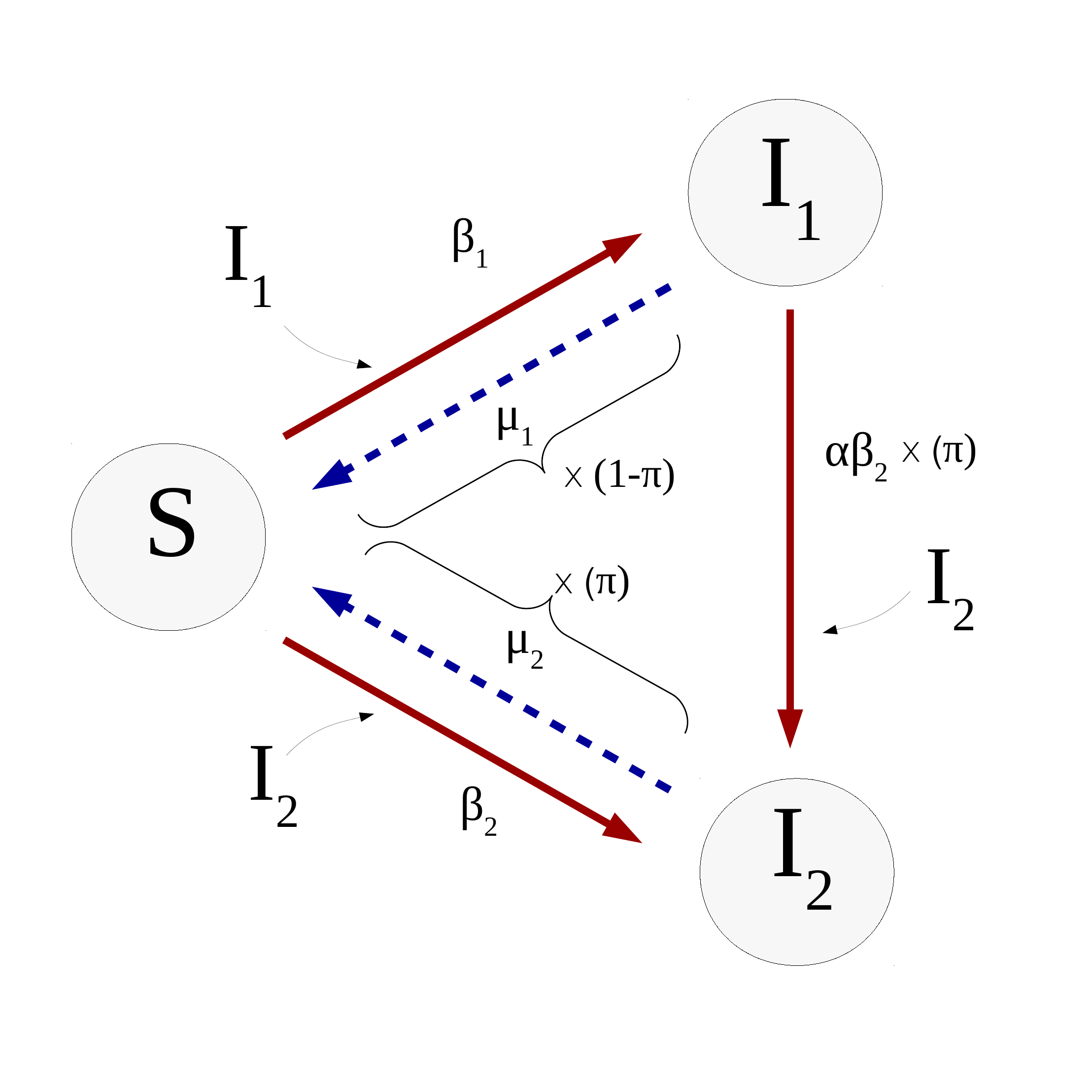}
    \caption{State transitions of model B. Infection and healing rates of disease I (II) are respectively $\beta_1$ ($\beta_2$) and $\mu_1$ ($\mu_2$). I$_1$ individuals can be (super)infected by disease II with rate $\alpha\beta_2$. Each rate is multiplied by its corresponding time scale factor: $1 - \pi$ for disease I processes and $\pi$ for disease II processes.}
    \label{fig:superinf_scheme}
    \end{center}
\end{figure}

Specifically, in model B, we represent susceptible individuals simply by S, and infected individuals of diseases I and II, respectively, by I$_1$ and I$_2$. The transmission rates are $\beta_1$ and $\beta_2$, the healing rates are $\mu_1$ and $\mu_2$, and the rate at which I$_1$ individuals are ``superinfected'' by disease II when exposed to I$_2$ individuals is given by a modified term $\alpha\cdot\beta_2$. As in model A, each term is also multiplied by the corresponding time scale factor ($(1 - \pi)$ for disease I and $\pi$ for disease II). The transitions are schematically represented in figure \ref{fig:superinf_scheme}. Despite the competitive aspect of the model, one can generate an asymmetrical interaction by setting $\alpha > 1$. This is because I$_1$ individuals, in this case, are more easily infected by disease II than susceptible ones, meaning that the spreading of disease II is enhanced by the presence of disease I, which in turn is in disadvantage due to the competition with disease II.

Using the same notation as before, $\rho_x$, for the density of individuals in state X, we write the dynamical equations:

\begin{eqnarray}
    \frac{d\rho_{s}}{dt} & = & -(1-\pi) \beta_1 \rho_{s}\rho_{i_1} - \pi \beta_2\rho_{s}\rho_{i_2} + (1-\pi)\mu_1\rho_{i_1} + \nonumber \\ && + \; \pi\mu_2\rho_{s} \nonumber\\
    \frac{d\rho_{i_1}}{dt} & = & (1-\pi) \beta_1 \rho_{s}\rho_{i_1} - \pi \alpha \beta_2 \rho_{i_1} \rho_{i_2} - (1 - \pi)\mu_1\rho_{i_1} \nonumber\\
    \frac{d\rho_{i_2}}{dt} & = & \pi \beta_2 \rho_{s}\rho_{i_2} + \pi \alpha \beta_2 \rho_{i_1} \rho_{i_2} - \pi \mu_2 \rho_{i_2},\nonumber
\end{eqnarray}
where the normalization constraint is $\rho_{s} + \rho_{i_1} + \rho_{i_2} = 1$. As in model A, we make the change of variables $u = \rho_{i_1}$, $v = \rho_{i_2}$ to obtain the reduced set of dynamical equations:

\begin{eqnarray}
    \label{eq:superinf_mf_u}
    \Dot{u} &=& (1-\pi) \; [  \beta_1(1 - u - v) - \mu_1 ] u  - \pi \alpha  \beta_2 u v \\
    \label{eq:superinf_mf_v}
    \Dot{v} &=& \pi [\;  \beta_2(1 - u - v) - \mu_2] v  + \pi \alpha  \beta_2 u \, v. 
\end{eqnarray}

\section{Results}

\subsection{Phase diagrams}

\begin{figure}[!tb]
    \begin{center}
    \includegraphics[width=0.95\linewidth]{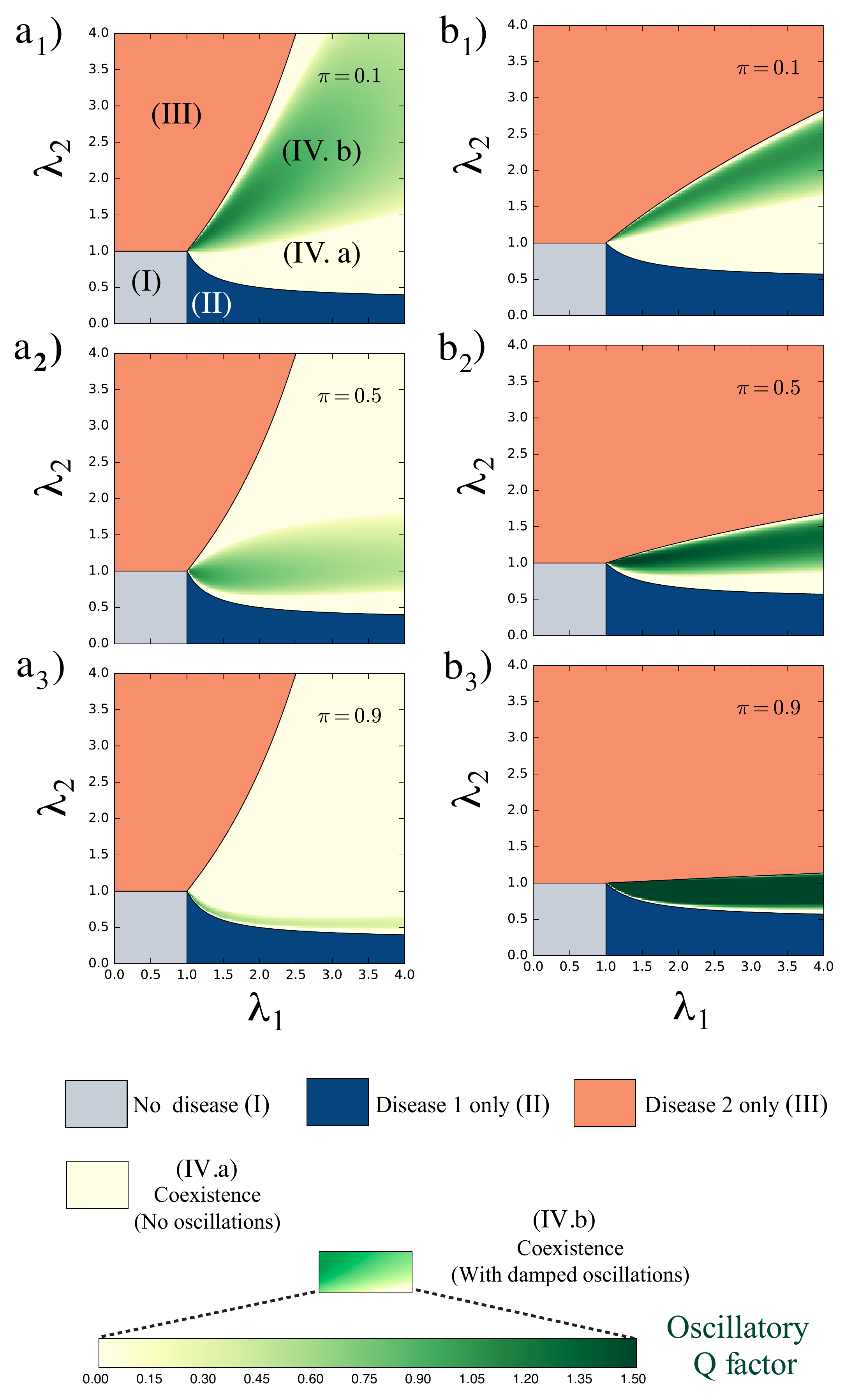}
    \caption{Phase diagrams of models A (a$_1$ to a$_3$) and B (b$_1$ to b$_3$), for three different values of the time scale parameter $\pi$. The phases are (I) no disease, (II) disease I only, (III) disease II only and (IV) coexistence. Inside the coexistence region, there is also the possibility of damped oscillations, quantified by the $Q$ factor defined in equation \ref{eq:q_factor} and exhibited here as a green scale. Other parameters are set to: $\Gamma_1 = 0.20$, $\Gamma_2 = 3.0$ and $\alpha = 2.0$.}
   \label{fig:phase_diags}
    \end{center}
\end{figure}

In the asymmetrically interacting regime, models A and B share a common feature: for any given set of parameters, there is exactly one stable fixed point within the region of the phase portrait that corresponds to physically possible solutions. Therefore, unlike what has been reported for mutually competitive or cooperative scenarios \cite{nowak1994superinfection,chen2017fundamental}, our models do not present bistability \footnote{For model B, it can be shown that the bistability condition given by Wu and collaborators in \cite{wu2013superinfection} cannot be met when $\alpha > 1$, which is our case.}. This in intrinsic to the positive-negative feedback of the asymmetric interaction between the diseases. Both models A and B present four phases, separated by transcritical bifurcations: (I) no disease, (II) disease I only, (III) disease II only and (IV) coexistence of both diseases. Here, we investigate the $\lambda_1 \times \lambda_2$ phase diagrams, setting the other parameters to fixed values. Figure \ref{fig:phase_diags} shows the phase diagrams for both models.

Through a stability analysis, we can obtain the phase transition curves of both models. The boundary between phases (II) and (IV) in model A is expressed as:

\begin{equation}
    \label{eq:2sis_threshold_lower}
    \frac{1}{\lambda_2} = 1 + (\Gamma_2 - 1)\left( 1 - \frac{1}{\lambda_1} \right), \quad \lambda_1 > 1,
\end{equation}
whereas, by the symmetry of the model, the boundary between phases (III) and (IV) is given by:

\begin{equation}
    \label{eq:2sis_threshold_upper}
    \frac{1}{\lambda_1} = 1 - (1 - \Gamma_1)\left( 1 - \frac{1}{\lambda_2} \right), \quad \lambda_2 > 1.
\end{equation}

The other two bifurcations are trivial and are given by $\lambda_2 = 1$ for $0 \leq \lambda_1 \leq 1$ (boundary between (I) and (III)) and $\lambda_1 = 1$ for $0 \leq \lambda_2 \leq 1$ ((I) and (II)). It is worth noticing that, for model A, none of the phase transition curves depends on the time scale parameter $\pi$. 

For model B, the boundary between (II) and (IV) is given by: 

\begin{equation}
    \label{eq:superinf_lower_curve}
    \frac{1}{\lambda_2} = 1 + (\alpha - 1)\left( 1 - \frac{1}{\lambda_1} \right), \quad \lambda_1 > 1,
\end{equation}
which is similar to that in equation \ref{eq:2sis_threshold_lower}, only replacing $\Gamma_2$ by $\alpha$. The boundary between (III) and (IV) is expressed as:

\begin{equation}
    \label{eq:superinf_upper_curve}
    \lambda_1 = \left[ \alpha \chi (\lambda_2 - 1) + 1  \right] \lambda_2, \quad \lambda_2 > 1,
\end{equation}
where we define $\chi$ (interpreted as the time scale ratio between diseases II and I) as:
\begin{equation}
    \label{eq:chi_def}
    \chi = \frac{\pi \mu_2}{(1 - \pi)\mu_1}.\nonumber
\end{equation}
The trivial boundaries of region (I) with regions (II) and (III) are the same as in model A.
Notice, however, that the parameter $\chi$ depends increasingly on the time scale parameter $\pi$, and so does the critical $\lambda_1$ value from equation \ref{eq:superinf_upper_curve}. This means that, as $\pi$ increases (i.e., when disease II propagates on shorter time scales), the phase transition between (III) and (IV) moves to lower values of $\lambda_2$, contracting the region of coexistence. Therefore, in this model, a faster clock for disease II makes it more effective to suppress disease I. This is an interesting result that might be used to control the prevalence of disease (or any other "infectious agent") in the host population. The $\pi$-dependence of the phase diagrams of model B constitutes an important difference with respect to model A, and was already reported by Wu and collaborators \cite{wu2013superinfection} as a dependency on the recovery rate when keeping the ratios $\lambda / \mu$ constant. 

\subsection{Behavior of the stationary prevalence}

In the phase of coexistence (region (IV) of the phase diagrams), both models present a single stable fixed point, for which $u, v \neq 0$. One can use the dynamical equations (eqs. \ref{eq:2sis_mf_u} to \ref{eq:2sis_mf_w} for model A and \ref{eq:superinf_mf_u} and \ref{eq:superinf_mf_v} for model B) to derive analytical expressions for the prevalences at the coexistence fixed point. The derivation and the final expressions are shown in appendix \ref{sec:app_coex_prevalences}. As expected, the stationary prevalence of each disease is a non-decreasing function of its reproduction ratio $\lambda = \beta / \mu$, when considering other parameters as fixed. However, the dependence of the prevalences with the relative time scale parameter $\pi$ is not trivial, and is different in each model.

\begin{figure}
    \centering
    \includegraphics[width=0.95\columnwidth]{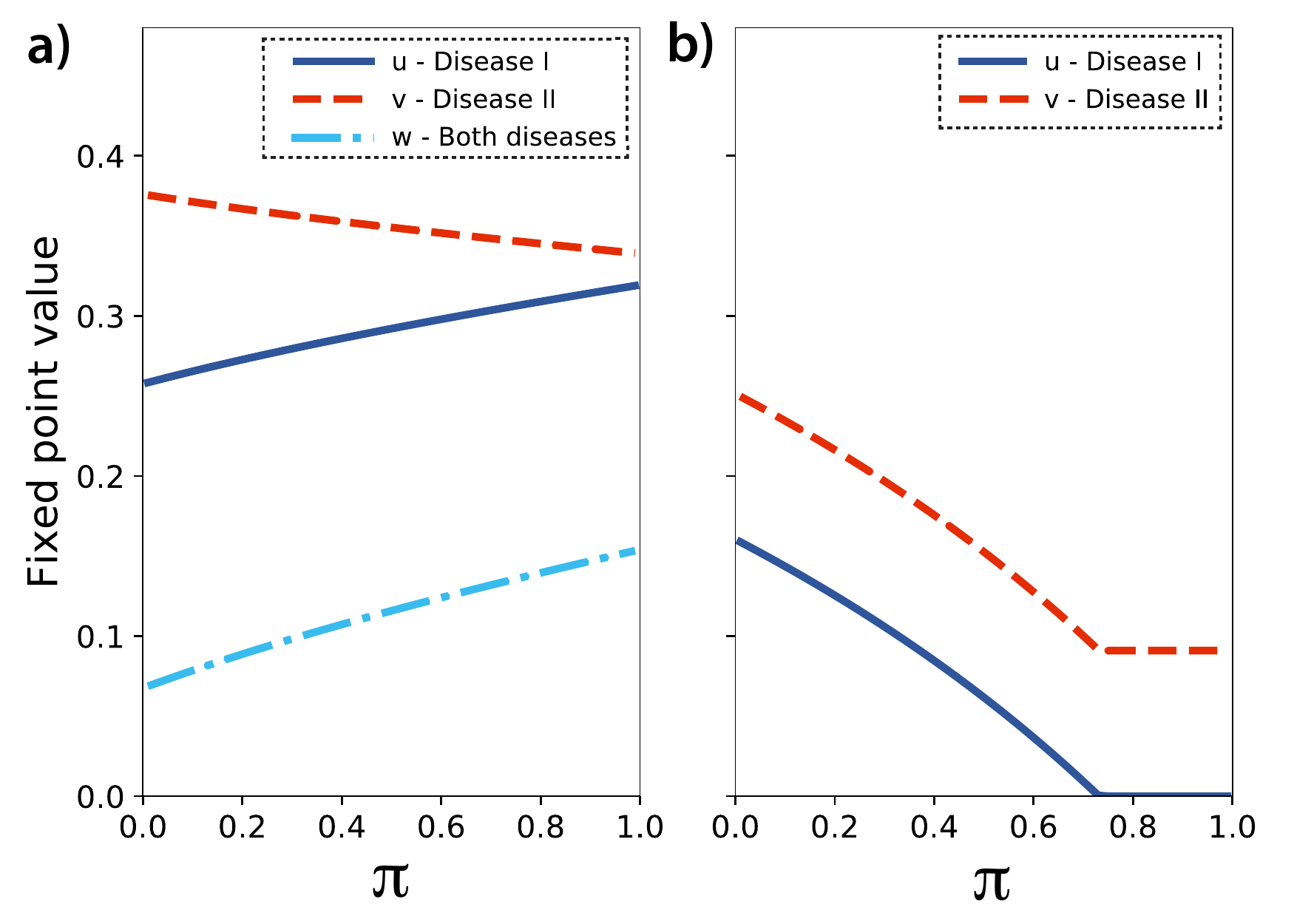}
    \caption{Stationary values of the prevalences as a function of $\pi$ for (a) model A and (b) model B. For both models, $\lambda_1$ and $\lambda_2$ are respectively set to 1.7 and 1.1. Other parameters are set to $\Gamma_1 = 0.5$, $\Gamma_2 = 2.5$ in (a) and $\alpha = 2.0$ in (b).}
    \label{fig:pi_prevalences}
\end{figure}

Figure \ref{fig:pi_prevalences} shows the basic behavior of the fixed point prevalences with $\pi$ for models A and B. While the prevalence $v$ of disease II decreases with $\pi$ for both models, the prevalence $u$ of disease I has opposite behaviors in each of them. In model A, the prevalence $w$ of coinfection increases with $\pi$. This variant shows the same behavior (if equating disease II with the information) as recently reported in some works on epidemic with awareness \cite{wang2017epidemic,da2019epidemic} in complex networks, namely, a faster relative clock of the information induces an increase on the disease prevalence. For model B, however, the behavior is the opposite: the prevalence of disease I decreases with $\pi$. Considering also how the time scale parameter distorts the phase diagram of model B (see figure \ref{fig:phase_diags}), we see that a faster clock of the ``predator'' process (disease II) effectively decreases (and possibly leads to the extinction of) the spreading of the ``prey'' process. Therefore, the relationship between the prevalences and the relative time scale in asymmetrically interacting spreading phenomena is a feature that depends on the specific shape of the considered model. 

\begin{figure}
    \centering
    \includegraphics[width=0.95\columnwidth]{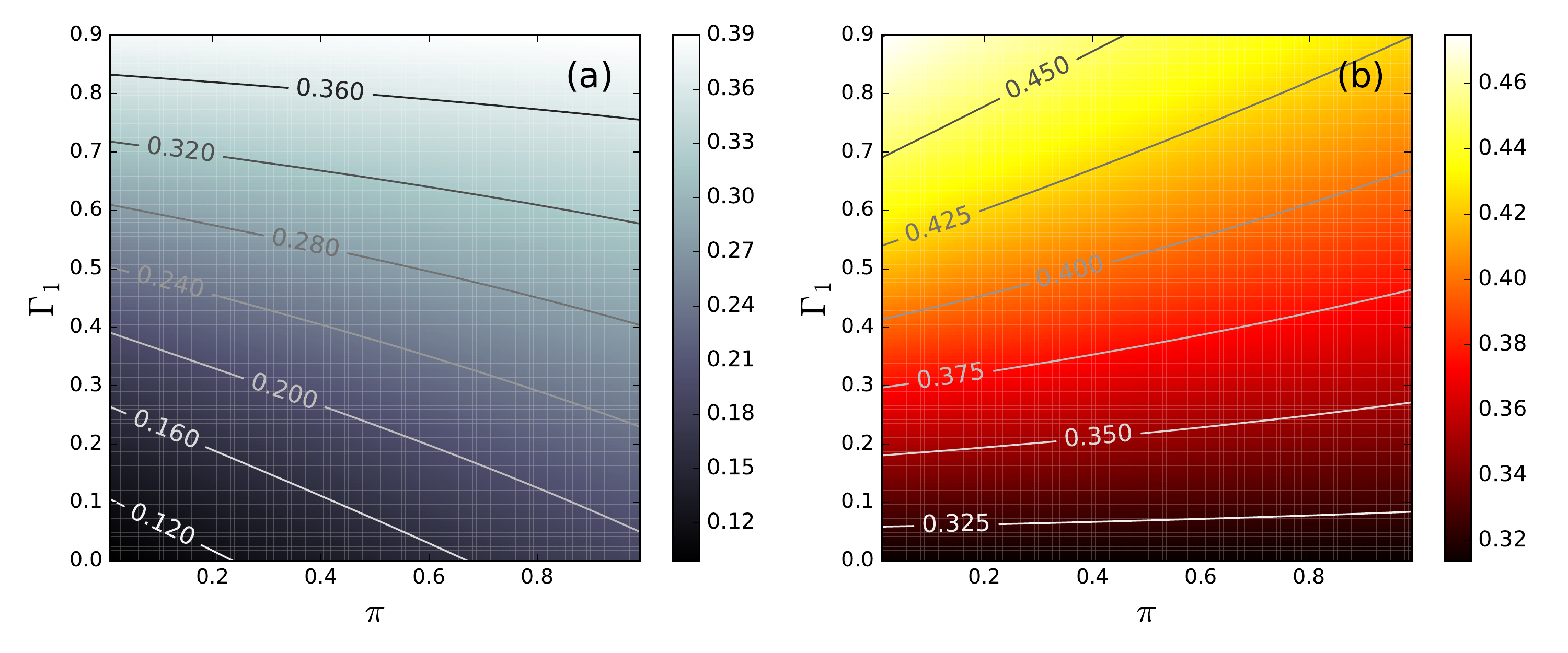}
    \includegraphics[width=0.95\columnwidth]{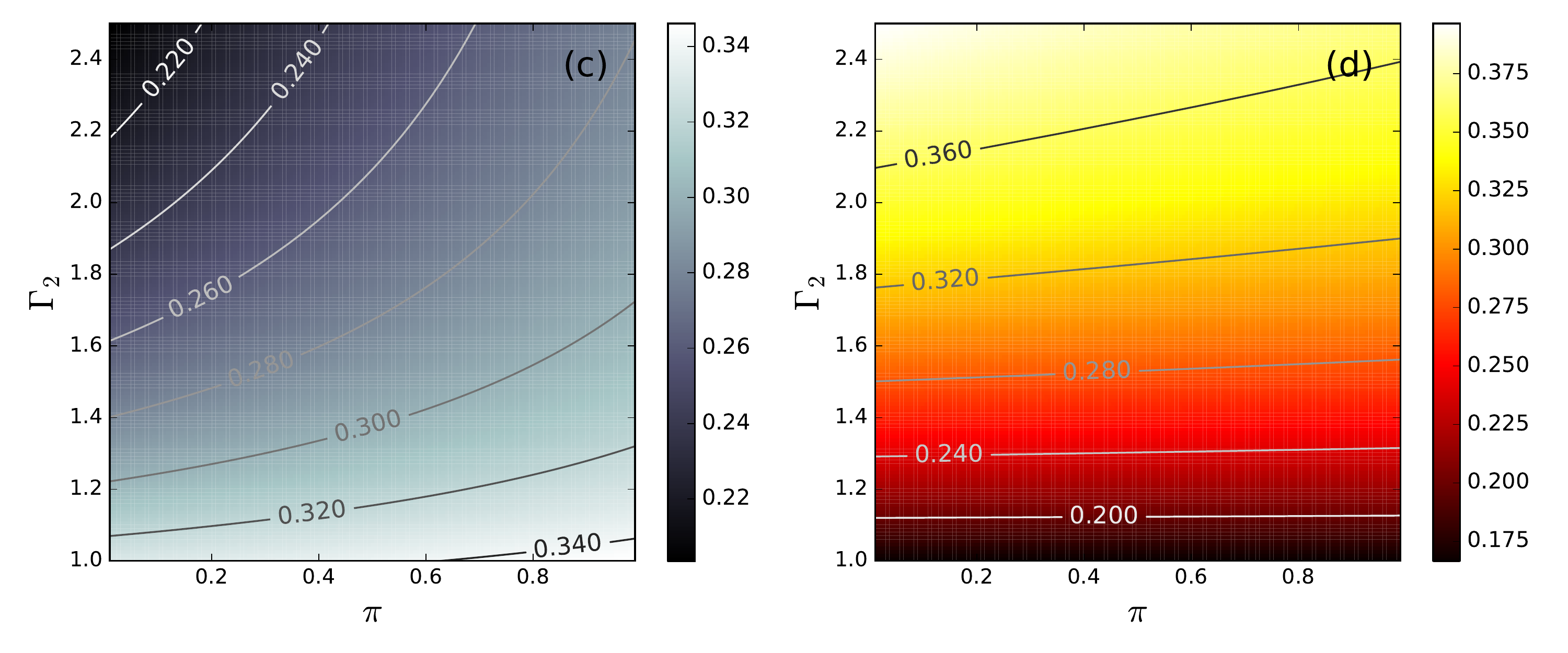}
    \caption{Values of the stationary prevalences of disease I ((a) and (c)) and disease II ((b) and (d)) for model A, plotted as functions of $\pi$, $\Gamma_1$ and $\Gamma_2$. In (a) and (b), $\Gamma_2$ is fixed to $2.5$, whereas in (c) and (d) $\Gamma_1$ is fixed to 0.4. The reproduction ratios are set to $\lambda_1 = 1.70$ and $\lambda_2 = 1.2$.}
    \label{fig:gamma_pi_diagrams}
\end{figure}

\begin{figure}
    \centering
    \includegraphics[width=0.95\columnwidth]{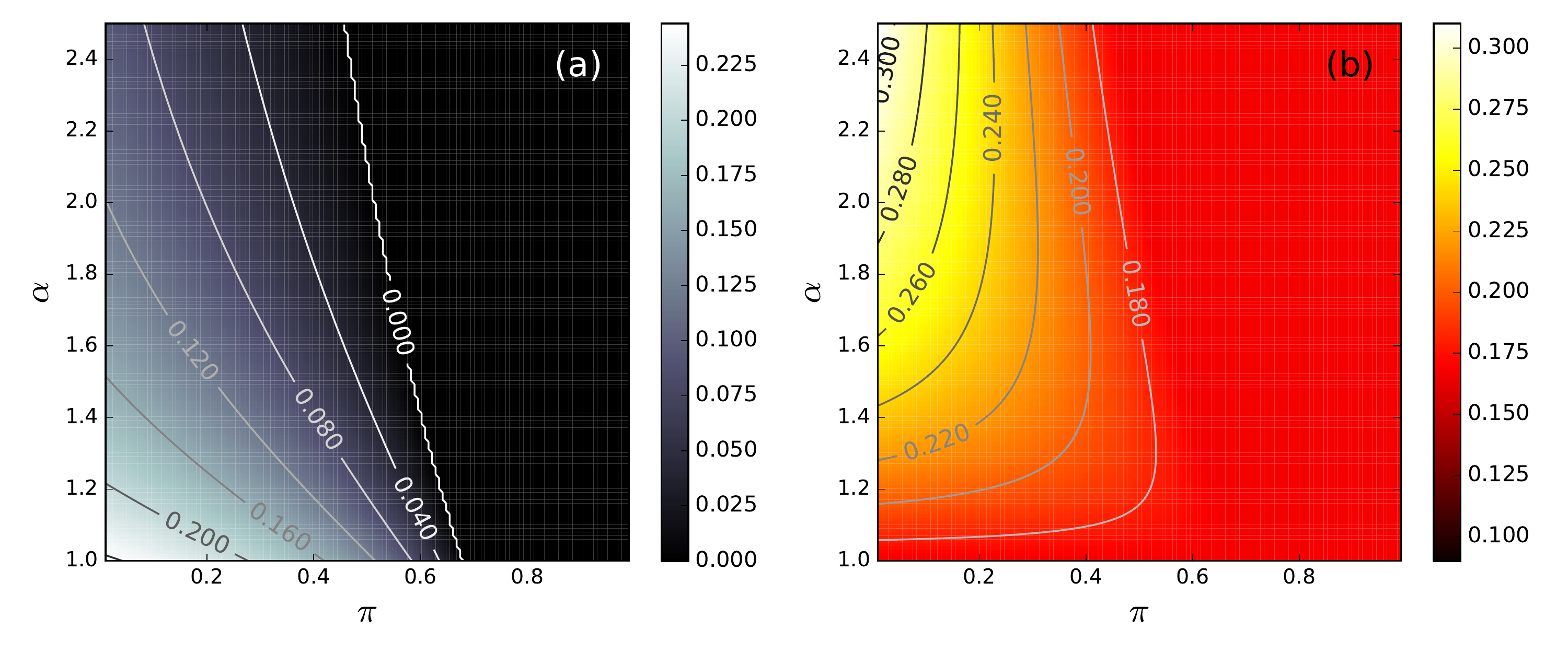}
    \caption{Stationary prevalences of (a) disease I and (b) disease II for model B, plotted as functions of $\pi$ and $\alpha$. The dashed line shows the position of the prevalence peaks. The reproduction ratios are set to $\lambda_1 = 1.7$ and $\lambda_2 = 1.2$.}
    \label{fig:alpha_pi_diagrams}
\end{figure}

In figures \ref{fig:gamma_pi_diagrams} and \ref{fig:alpha_pi_diagrams}, we further analyze the behavior of the prevalences with $\pi$ and the parameters that control the interactions: $\Gamma_1$, $\Gamma_2$ for model A and $\alpha$ for model B. We first notice that the increasing or decreasing trends with respect to $\pi$, as observed in figure \ref{fig:pi_prevalences}, are not changed for different values of the interaction parameters: disease I increases while disease II decreases with $\pi$ for model A (figure \ref{fig:gamma_pi_diagrams}), whereas both prevalences decrease with $\pi$ for model B (figure \ref{fig:alpha_pi_diagrams}). To rule out the possibility that there is a region of the parameter space in which the reported behaviors with respect to $\pi$ might be different, we show in appendix \ref{sec:app_behavior_with_pi} that this is not possible, i.e., that the behavior with $\pi$ is always the same for each model in the coexistence region.

We can also analyze how the prevalences vary with changes in the interaction parameters. For model A (figure \ref{fig:gamma_pi_diagrams}), we see that an increase in $\Gamma_1$ increases both the prevalences of disease I (a) and II (b). This is expected, as a greater $\Gamma_1$ value means a weaker impairing to the propagation of disease I, which is beneficial for both diseases (as disease II benefits from disease I). On the other hand, an increase in $\Gamma_2$ causes a decrease in disease I (c) and an increase in disease II (d). This is also expected, as a larger value of $\Gamma_2$ means a greater benefit to disease II, which in turn is detrimental to disease I. Therefore, for model A, the effect of the two interaction parameters in each disease's prevalence is intuitive and predictable. The effects are also numerically influenced by the time scale $\pi$, yet not qualitatively changed.

\begin{figure}
    \centering
    \includegraphics[width=0.95\columnwidth]{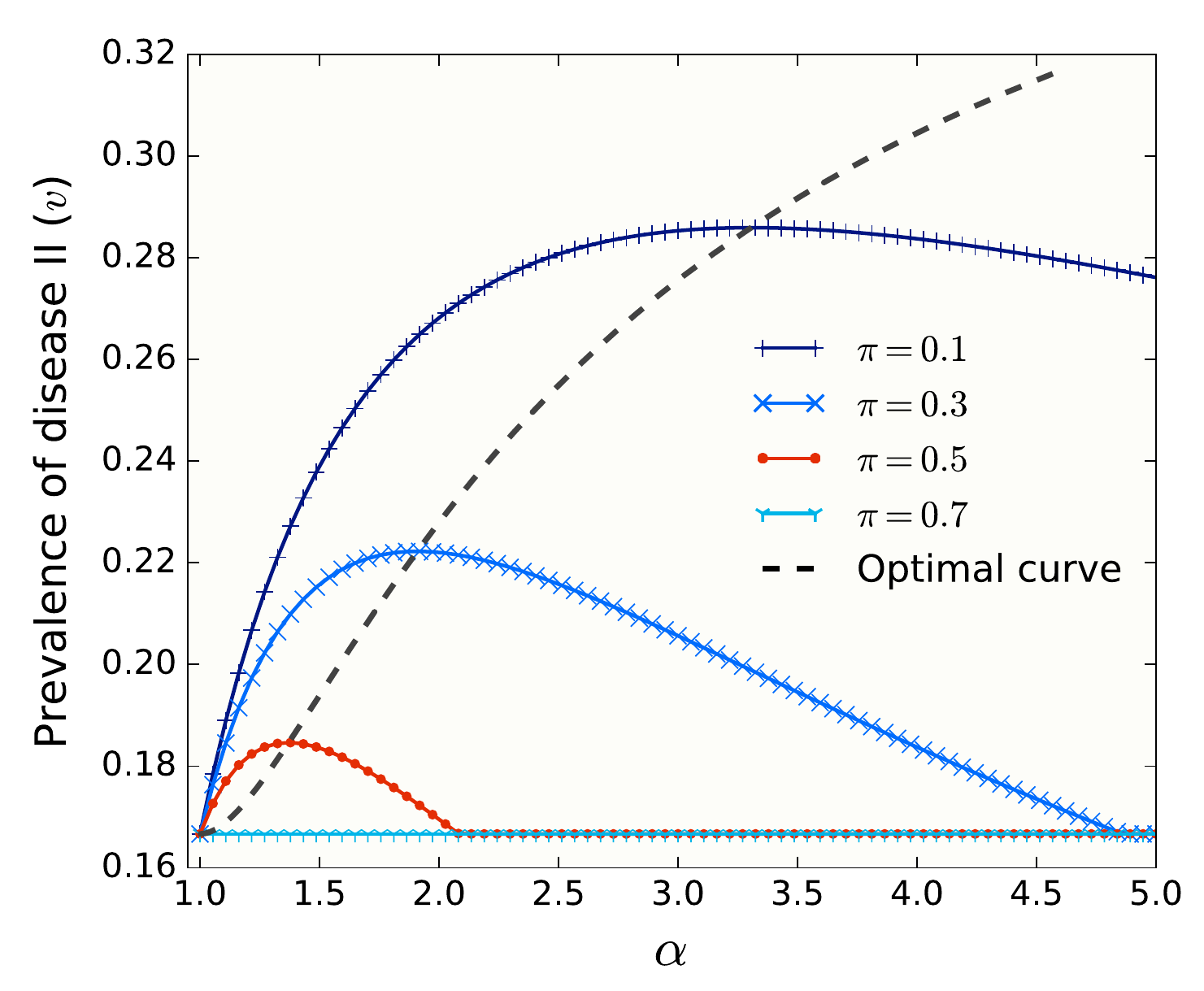}
    \caption{Stationary prevalences of disease II for model B, plotted as a function of and $\alpha$ and for different values of $\pi$. The dashed line represents the optimal value of $\alpha$ and its corresponding $v$ as $\pi$ is continuously changed. The other parameters are set to $\beta_1 = 1.7$, $\beta_2 = 1.2$ and $\mu_1 = \mu_2 = 1$.
    }
    \label{fig:v_of_alpha}
\end{figure}

However, for model B, which has a single interaction parameter $\alpha$, the behavior of the prevalence is not trivial. From figure \ref{fig:alpha_pi_diagrams}, we see that, while disease I prevalence (a) is always reduced with an increase in $\alpha$, the behavior of the prevalence of disease II (b) with $\alpha$ is not uniform, and may have an optimal value that depends on $\pi$. This happens because the superinfection transition, controlled by $\alpha$, is simultaneously beneficial to disease II and detrimental to disease I. Thus, as seen from model A, an increase in $\alpha$ certainly reduces the prevalence of disease I, but has a ``conflicting'' effect to the prevalence of disease II. While a value of $\alpha$ close to 1 means almost no benefit to disease II from disease I, a large value $\alpha \gg 1$ means an excessive ``predation'' from disease II, therefore existing an optimal intensity of the interaction $\alpha$. This is further illustrated in Figure \ref{fig:v_of_alpha},where we show the prevalence of disease II in model B as a function of $\alpha$, for different values of $\pi$. As it can be seen, there is an optimal value of $\alpha$ that maximizes the prevalence, for fixed values of the other parameters.

\subsection{Damped oscillations: node \emph{vs} spiral point}

Some predator-prey systems, which naturally have an asymmetrical relationship between two processes, are known to present stable closed orbits (i.e., sustained oscillations) \cite{brauer2001mathematical}. For both epidemic models addressed in our work, there are no closed orbits in the physical region of the phase portrait \footnote{This can be proven, for model B, using Dulac's criterion, as done in this book \cite{brauer2001mathematical}.}. However, in the coexistence phase (region (IV)), the stable fixed point can be either a node or a spiral point. In the second case, the transient dynamics of the system towards the fixed point may present some damped oscillations. The presence of such local oscillations is determined by the imaginary part of the eigenvalues of the model's Jacobian matrix, calculated at the stable fixed point. For model B, which is two-dimensional, the Jacobian's eigenvalues $\sigma_1, \sigma_2$ can either be both real or complex conjugate to each other. For model A, which is three-dimensional, the 3x3 Jacobian matrix can either have none or two non-real conjugate eigenvalues. For both models A and B, one can use the quantity:

\begin{equation}
    \label{eq:q_factor}
    Q = \max_{i = 1, \;..., d}\left(  \left|\frac{\text{Im}(\sigma_i)}{\text{Re}(\sigma_i)} \right| \right),
\end{equation}
to measure the ``quality factor'' of the oscillations when $\text{Re}(\sigma_i) \neq 0$ ($\{ \sigma_i \}$ are the eigenvalues of the Jacobian at the fixed point, $d$ is the dimensionality of the system). This is because the imaginary part is responsible of the oscillations, and the real part of the damping, thus the imaginary-to-real part ratio measures the propensity of the system to oscillate around the fixed point.

\begin{figure}[t]
    \centering
    \includegraphics[width=0.99\columnwidth]{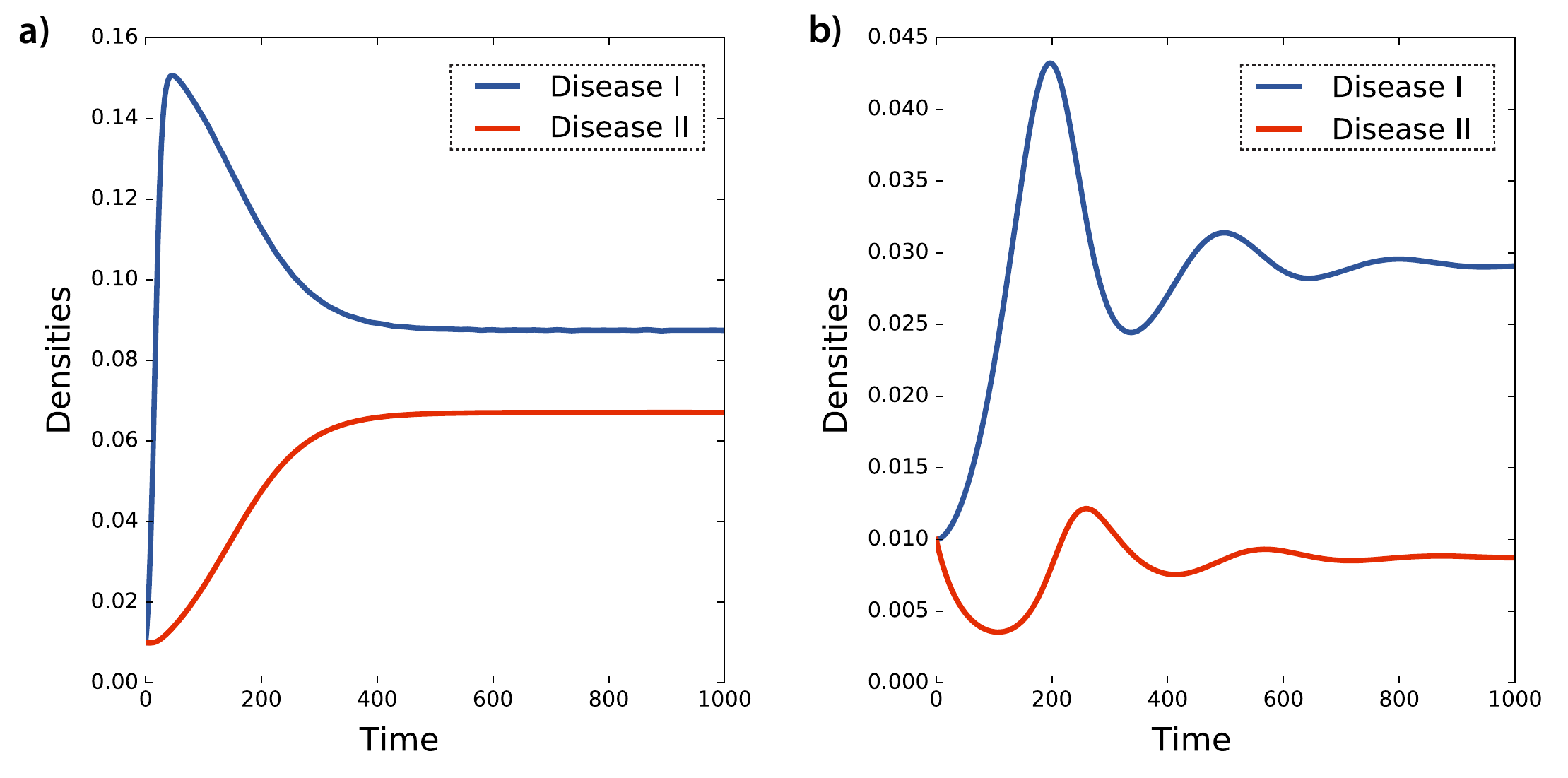}
    \caption{Time evolution of model B prevalences for two different conditions: (left) $\pi = 0.1$, for which $Q = 0$, meaning that there are no local oscillations around the fixed point, and (right) $\pi = 0.9$, for which $Q = 3.65$ and the system oscillates before converging to the steady state. Other parameters are set to: $\beta_1 = 1.20$, $\beta_2 = 0.98$, $\mu_1 = \mu_2 = 1.0$ and $\alpha = 2.0$. The initial fraction of infected individuals is set to $0.01$ for both diseases.}
    \label{fig:sample_oscillations_superinf}
\end{figure}

In Figure \ref{fig:phase_diags}, together with the regular phases of the model, we show in (color-coded) green the numerically calculated values of Q. The coexistence phase can thus be subdivided according to the existence of a non-real Jacobian eigenvalue: within the white regions, the eigenvalues are real and the fixed point is a node whereas in the green areas, the fixed point is a spiral point and there may occur oscillations around it. Notice that, for both models A and B, the shape of the spiral point region depends considerably on the time scale parameter $\pi$. Greater values of $\pi$ seem to shrink the oscillatory region and reduce the values of Q for model A, but the opposite appears to happen with model B. Furthermore, we show the difference between a node and a spiral point in figure \ref{fig:sample_oscillations_superinf}, in which the time evolution of the prevalences of model B is shown for two situations: one with $Q = 0$ (hence no local oscillations (left)), and another with $Q = 3.65$ (thus there are damped oscillations before reaching the steady state). Interestingly, we only had to change the relative time scale parameter $\pi$ to switch between the two situations.

An important fact is that damped oscillations can only occur when the interaction between the diseases is asymmetrical, for both models A and B. For model A, we numerically check this by observing that the oscillatory region of the phase diagram shrinks and disappears as $\Gamma_1$ or $\Gamma_2$ leaves the region for which interactions are asymmetric. For model B, it can be shown that the Jacobian's eigenvalue equation (which is quadratic) can only assume non-real solutions if $\alpha > 1$. Finally, we note that despite that our deterministic formulation predicts that the oscillations are always damped, stochasticity - which is intrinsic to real-world systems, and are inherent when performing Monte Carlo simulations - could cause such oscillations to last for the long term, as small perturbations to the prevalences could recover the oscillatory pattern.

\section{Conclusions}

In this work, we have studied two minimalist models for interacting diseases in the asymmetrical regime, for homogeneously mixed populations and with continuous-time evolution. We focus on the influence of the relative time scale between the two diseases. The simplicity of our framework not only provides us with analytical tractability, but also reveals the fundamental properties of asymmetrically interacting contagion. The models are simple enough to be applied to different situations, and the choice for two models is justified by our goal of achieving a more general understanding about interacting processes.

Model A is a mathematical prototype for epidemic with awareness and the possible asymmetrical interaction between HIV and some specific diseases \cite{watt2003decrease,moss2002suppression,xiang2009viruses}. Our results show that, despite the epidemic thresholds are not affected by the time scale parameter $\pi$, the prevalence is influenced in a non-intuitive way: the ``prey'' disease (I) has greater prevalence when the ``predator'' disease (II) has a faster time scale. If we take the HIV as an example, which naturally has a long time scale for its development, detection and treatment, this means that the information awareness (which is more quickly transmitted and forgotten) may not be as efficient as it could due to its shorter time scale. 

Model B, which is inspired in situations such as computer viruses and spreading countermeasures \cite{chen2004impact,goldenberg2005distributive,zhu2013mixing,yang2014pulse} or fake \emph{vs} fact-checked news, has a different behavior with the relative time scale: if the ``predator'' disease (II) has a faster clock, both prevalences of diseases I and II decay, with the possibility of disease I eradication (see figure \ref{fig:pi_prevalences}.b). The interpretation then varies according to the situation being modeled. For example, if the main ``goal'' of disease II is to fight the other disease (such as a fact-checked against fake news), then a shorter time scale (higher $\pi$) is desirable. If however the goal is to maximize disease II prevalence by taking advantage of disease I (one can imagine a computer virus that makes use of vulnerabilities generated by another one), then a longer time scale for disease II is preferable. 

Another interesting phenomenology revealed is the behavior of model B with its superinfection parameter $\alpha$, which controls the interaction between diseases I and II in both directions. The interpretation again depends on the situation: if the goal is to minimize disease I, then a greater value of $\alpha$ is always better, whereas for the optimization of the prevalence of disease II an intermediate value of $\alpha$ (which depends on $\pi$) should be sought. Finally, we have shown that oscillatory behavior, which is a well-known feature of predator-prey systems, is also present in asymmetrically interacting epidemic models, and that its expression crucially depends on the relative time scale. Although the theory predicts that they are strongly damped, real world systems could persistently display such oscillations due to random fluctuations. 

To round off, we note that apart from the above connections to real dynamical systems, we expect that our work helps future studies of asymmetrically interacting spreading phenomena by providing general guidelines. By studying two different models, we provide a more general understanding of such systems, but we recognize that the results may have some bias due to the specific choice of models. An interesting extension of our work would be to consider a broader family of models, using more generic formulations, and determine for example the conditions that make the prevalences to depend, in one way or another, on the time scale. Finally, we also expect some variations when these models are studied on top of structured populations. For such scenarios, our results provide a baseline of which qualitative and quantitative patterns are not to be associated to a network effect. 

\section{Acknowledgments}

PCV thank Fapesp for the financial support through grants no. 2016/24555-0 and 2019/11183-5. Research carried out using the computationalresources of the Center for Mathematical Sciences Applied to Industry (CeMEAI) funded by FAPESP (grant 2013/07375-0). FAR acknowledges CNPq (grant 307974/2013-8) and FAPESP (grants 2017/50144-0 and 16/25682-5) for the financial support given for his research. YM acknowledges partial support from the Government of Aragón, Spain through grant E36-20R, by MINECO and FEDER funds (grant FIS2017-87519-P) and by Intesa Sanpaolo Innovation Center. 

\appendix

\section{Analytical expressions for the coexistence fixed point}
\label{sec:app_coex_prevalences}

The coexistence phase is characterized by the long-term presence of both diseases, thus having $u, v > 0$ in the steady state. In this appendix, we show how to analytically calculate the prevalences for the coexistence fixed points of both models. For simplicity of notation, throughout the whole appendix section we use $u$, $v$ and $w$ to denote the fixed point values of the prevalences (and not their time-dependent functions). 

\subsection{Model A}

We find the fixed points by setting $\dot{u} = \dot{v} = \dot{w} = 0$ in the dynamical equations \ref{eq:2sis_mf_u} to \ref{eq:2sis_mf_w}. Using the fact that $u, v > 0$ at the coexistence, we divide equation \ref{eq:2sis_mf_u} by $(1-\pi)\mu_1u$, equation \ref{eq:2sis_mf_v} by $\pi\mu_2v$ and equation \ref{eq:2sis_mf_w} by $(1-\pi)\mu_1$, obtaining the following reduced system of equations:
\begin{eqnarray}
    \label{eq:2sis_reduced_u}
    0 &=& \lambda_1[1 - u + (\Gamma_1 - 1)(v-w)] - 1 \\
    \label{eq:2sis_reduced_v}
    0 &=& \lambda_2[1 - v + (\Gamma_2 - 1)(u-w)] - 1 \\
    \label{eq:2sis_reduced_w}
    0 &=& \lambda_1\Gamma_1(v-w)u - w + \chi[\lambda_2\Gamma_2(u-w)v - w]
\end{eqnarray}
in which only equation \ref{eq:2sis_reduced_w} is non-linear. The disease II time scale factor $\chi$ is the same as defined in Eq. \ref{eq:chi_def}. Using also the definition:
\begin{equation}
    \label{eq:simple_sis_def}
    s_i = 1 - \frac{1}{\lambda_i}, \quad i = 1, 2,
\end{equation}
where $s_i$ is the solution for a non-interacting SIS model with reproduction ratio $\lambda_i$, we can further simplify equations \ref{eq:2sis_reduced_u} and \ref{eq:2sis_reduced_v} respectively to:
\begin{eqnarray}
    \label{eq:2sis_us1}
    u - (\Gamma_1 - 1)(v - w) = s_1 \\
    \label{eq:2sis_vs2}
    v - (\Gamma_2 - 1)(u - w) = s_2,
\end{eqnarray}
from which it is intuitive to see that the interaction between the diseases (represented by $\Gamma_1$ and $\Gamma_2$) causes a deviation from the non-interacting solution.

At this point, we split the solution in two cases: (a) the simpler case $\Gamma_1 = 0$ (i.e., when disease II completely inhibits infection by disease I) and (b) the more general case $\Gamma_1 > 0$.

\subsubsection{Case $\Gamma_1 = 0$}

In this case, equation \ref{eq:2sis_us1} simplifies to:

\begin{equation}
    u - w = s_1 - v
\end{equation}

Combining this equation with \ref{eq:2sis_vs2}, we get a 2x2 system for $v$ and $(u-w)$, for which the solution is:

\begin{eqnarray}
    \label{eq:2sis_v_partialsol}
    v = \frac{s_2 + (\Gamma_2 - 1)s_1}{\Gamma_2} \\
    \label{eq:2sis_umw_partialsol}
    u - w = \frac{s_1 - s_2}{\Gamma_2}
\end{eqnarray}

Now using that $\Gamma_1 = 0$ in equation \ref{eq:2sis_reduced_w}, we obtain an expression for $w$ in terms of known variables:

\begin{equation}
    w = \frac{\chi}{1 + \chi} \lambda_2 \Gamma_2 (u-w) v
\end{equation}

Plugging equations \ref{eq:2sis_v_partialsol} and \ref{eq:2sis_umw_partialsol} in the above equation, and then using it back to equation \ref{eq:2sis_umw_partialsol}, we get the full expressions for the fixed points when $\Gamma_1 = 0$:

\begin{eqnarray}
    u &=& s_1 \! - \! \left[ 1 \! -\! \frac{\chi}{1 + \chi} \frac{s_1 - s_2}{1 - s_2} \right] \!\! \left[ \frac{s_2 + (\Gamma_2 - 1)s_1}{\Gamma_2} \right] \\
    v &=& \frac{s_2 + (\Gamma_2 - 1)s_1}{\Gamma_2} \\
    w &=& \frac{\chi}{1 + \chi} \frac{s_1 - s_2}{1 - s_2} \left[ \frac{s_2 + (\Gamma_2 - 1)s_1}{\Gamma_2} \right],
\end{eqnarray}
where we have also replaced $\lambda_2 = 1 / (1 - s_2)$. With some analysis of the above expressions, noticing that $\chi / (1 - \chi)$ is an increasing function of $\chi > 0$, we can infer that $u$ and $w$ increase with $\chi$ (and thus with $\pi$), whereas $v$ does not depend on the time scale factor. This is in agreement with the plots in figure \ref{fig:gamma_pi_diagrams} (a) and (b).

\subsubsection{Case $\Gamma_1 > 0$}

In this case, the fact that equation \ref{eq:2sis_reduced_w} is quadratic on its variables cannot be avoided. Our strategy is to use equations \ref{eq:2sis_reduced_u} and \ref{eq:2sis_reduced_v} to write $u$, $v$, $(u-w)$ and $(v-w)$ as functions of the variable $w$ and the model parameters. This can be used in equation \ref{eq:2sis_reduced_w} to find the solution for $w$ and, consequently, for the other variables.

We can isolate $v$ in equation \ref{eq:2sis_vs2} and apply it to equation \ref{eq:2sis_us1}, obtaining:

\begin{eqnarray}
    \label{eq:u_of_w}
    u &=& m[P_{12} - (\Gamma_1 - 1)\Gamma_2 w] \\
    \label{eq:v_of_w}
    v &=& m[P_{21} - (\Gamma_2 - 1)\Gamma_1 w],
\end{eqnarray}
where we simplified the notation using the definitions

\begin{eqnarray}
    P_{12} &=& s_1 + (\Gamma_1 - 1)s_2 \\
    P_{21} &=& s_2 + (\Gamma_2 - 1)s_1 \\
    m &=& 1 / [1 - (\Gamma_1 - 1)(\Gamma_2 - 1)].
\end{eqnarray}

We can also manipulate equations \ref{eq:u_of_w} and \ref{eq:v_of_w} to obtain:

\begin{eqnarray}
    \label{eq:umw_of_w}
    u - w &=& m[P_{12} - \Gamma_1 w] \\
    \label{eq:vmw_of_w}
    v - w &=& m[P_{21} - \Gamma_2 w]
\end{eqnarray}

Now we plug the above expressions into equation \ref{eq:2sis_reduced_w} which, after redistributing the terms and dividing them by $\Gamma_1 \Gamma_2 m^2$ ($m>0$ for asymmetrical interactions and, by hypothesis, $\Gamma_1 > 0$), becomes the quadratic equation $aw^2 + bw + c = 0$, where:

\begin{eqnarray}
    \label{eq:2sis_aw_coeff}
    & & a = \lambda_1 (\Gamma_1 - 1) \Gamma_2 +  \chi \lambda_2 (\Gamma_2 -1) \Gamma_1 \\
    \label{eq:2sis_bw_coeff}
    \nonumber \lefteqn{b = - \Big{\{} \lambda_1[ P_{12} + (\Gamma_1 - 1)P_{21} ] + }  \\ 
    & & + \chi\lambda_2[ P_{21} + (\Gamma_2 - 1)P_{12} ]  + \frac{1+\chi}{m^2\Gamma_1\Gamma_2} \Big{\}} \\
    & & c = \frac{P_{12}P_{21}}{\Gamma_1\Gamma_2} (\Gamma_1 \lambda_1 + \chi \Gamma_2\lambda_2)
\end{eqnarray}

For asymmetrical interactions, it is possible that $a = 0$ in the coexistence region. Thus we write the coexistence fixed point of the system as:

\begin{eqnarray}
    \label{eq:2sis_w_generalsol}
    w &=& 
    \begin{cases}
        \frac{-b - \sqrt{b^2 - 4ac}}{2a} & a \neq 0 \\ 
        \frac{-c}{b} & a = 0
    \end{cases}
    \\
    \label{eq:2sis_u_generalsol}
    u &=& m[P_{12} - (\Gamma_1 - 1)\Gamma_2 w] \\
    \label{eq:2sis_v_generalsol}
    v &=& m[P_{21} - (\Gamma_2 - 1)\Gamma_1 w]
\end{eqnarray}

From the above expressions, it is difficult to extract the behavior of the prevalences with respect to the time scale parameter $\chi$ (or $\pi$). However, in appendix \ref{sec:app_behavior_with_pi} we present an alternative argument that reinforces the behavior observed in Fig. \ref{fig:gamma_pi_diagrams}.

\subsection{Model B}

Model B has a much simpler procedure to find analytical expressions for the coexistence fixed point $u$ and $v$, for general values of the parameters. Knowing that $u, v > 0$, one can divide eq. \ref{eq:superinf_mf_u} by $(1-\pi)\mu_1u$ and eq. \ref{eq:superinf_mf_v} by $\pi\mu_2v$ and set their left-hand sides to $0$, obtaining:

\begin{eqnarray}
    0 &=& \lambda_1 (1 - u - v) - 1 - \chi \alpha \lambda_2 v \\
    0 &=& \lambda_2(1 - u - v) - 1 + \alpha \lambda_2 u
\end{eqnarray}
which is a 2x2 linear system in $u$ and $v$. Passing convenient terms to the left side of each equation, one can write the system in terms of $s_1$ and $s_2$ as defined in \ref{eq:simple_sis_def}:

\begin{eqnarray}
    s_1 &=& u + \phi v \\
    s_2 &=& (1-\alpha)u + v
\end{eqnarray}
where we define $\phi$ as:

\begin{equation}
    \phi = 1 + \alpha \chi \lambda_2 / \lambda_1 
\end{equation}

The solution of this 2x2 system is:

\begin{eqnarray}
    \label{eq:solution_superinf_u}
    u & = & \frac{s_1 - \phi s_2}{1 + \phi (\alpha - 1)} \\
    \label{eq:solution_superinf_v}
    v & = & \frac{(\alpha - 1)s_1 + s_2}{1 + \phi (\alpha - 1)}
\end{eqnarray}

\section{Behavior of the prevalences with $\pi$}
\label{sec:app_behavior_with_pi}

By plotting the values of the fixed point prevalences, analytically derived in appendix \ref{sec:app_coex_prevalences}, we could analyze the behavior of such prevalences with the time scale parameter $\pi$ for models A and B. In this section, we provide analytical support for the observed behaviors in both models.

\subsection{Model A}

A possible approach to determine the slope of the fixed point prevalences $u, v, w$ with $\pi$ is to differentiate the expressions \ref{eq:2sis_w_generalsol} to \ref{eq:2sis_v_generalsol} with respect to $\chi$ (notice, from the definition in equation \ref{eq:chi_def}, that $\chi$ is an increasing function of $\pi$). This procedure, however, can be very tedious and provides little or no analytical insight. An alternative approach is to implicitly differentiate the reduced equations \ref{eq:2sis_reduced_u} to \ref{eq:2sis_reduced_w} w.r.t. $\chi$, obtaining more insightful expressions.

Let us define here $u' = \partial u / \partial \chi$, $v' = \partial v / \partial \chi$ and $w' = \partial w / \partial \chi$. Implicit differentiation of equations \ref{eq:2sis_reduced_u} to \ref{eq:2sis_reduced_w} yield:

\begin{eqnarray}
    0 &=& -\lambda_1 u' + (\Gamma_1 - 1) \lambda_1 v' - (\Gamma_1 - 1)\lambda_1 w' \\
    0 &=& -\lambda_2 v' + (\Gamma_2 - 1) \lambda_2 u' - (\Gamma_2 - 1)\lambda_2 w' \\ \nonumber
    0 &=& \Gamma_1 \lambda_1 (v' - w') + \Gamma_1 \lambda_1 (v - w) u' - w' +  \\ \nonumber && [ \Gamma_2 \lambda_2 (u - w)v - w ] + \\ && \chi [ \Gamma_2 \lambda_2 (u' - w')v + \Gamma_2 \lambda_2 (u - w) v' - w' ]
\end{eqnarray}

With some rearrangement, the above expressions can be written as a linear system given by:

\begin{equation}
    A \overrightarrow{x'} = \overrightarrow{b}
\end{equation}
where $\overrightarrow{x'} = (u', v', w')^T$, and

\begin{equation}
    A = 
    \begin{bmatrix}
        -\lambda_1 & (\Gamma_1 - 1)\lambda_1 & - (\Gamma_1 - 1)\lambda_1 \\
        (\Gamma_2 - 1)\lambda_2 & -\lambda_2 & -(\Gamma_2 - 1)\lambda_2 \\
        A_{wu} & A_{wv} & A_{ww}
    \end{bmatrix} 
\end{equation}
with

\begin{eqnarray}
    A_{wu} &=& \Gamma_1\lambda_1(v - w) + \chi\Gamma_2\lambda_2 v \\
    A_{wv} &=& \chi\Gamma_2\lambda_2(u - w) + \Gamma_1 \lambda_1 u \\
    A_{ww} &=& \Gamma_1\lambda_1 u + \chi \Gamma_2\lambda_2 v + 1 + \chi
\end{eqnarray}

Moreover, the vector of independent coefficients is

\begin{equation}
    \overrightarrow{b} = (0, 0, -[\Gamma_2\lambda_2(u-w)v - w])^T = (0, 0, b_w)^T
\end{equation}

Using Cramer's rule, we can obtain the $\chi$-derivatives as functions of the model parameters and the prevalences:

\begin{eqnarray}
    u' &=& (b_w \lambda_1 \lambda_2 / \det A) \; (1 - \Gamma_1) \Gamma_2 \\
    v' &=& (b_w \lambda_1 \lambda_2 / \det A) \; (1 - \Gamma_2) \Gamma_1 \\
    w' &=& (b_w \lambda_1 \lambda_2 / \det A) \; [1 - (\Gamma_1 - 1)(\Gamma_2 - 1) ] 
\end{eqnarray}

Thus, in the asymmetrically interacting regime ($0 \leq \Gamma_1 < 1$ and $\Gamma_2 > 1$), $u'$ has the same sign as $w'$ (thus $u$ and $w$ have the same slope with $\chi$ and $\pi$), whereas $v'$ ($v$) has opposite sign (slope). By showing that the ratio $b_w / \det A$ is positive, we could demonstrate that $u$ and $w$ actually increase with $\pi$, while $v$ decreases. Although we could not mathematically determine the signals of $b_w$ and $\det A$ for arbitrary model parameters, we collected robust numerical evidence that $b_w$ and $\det A$ are both negative in the coexistence phase (region IV in figure \ref{fig:phase_diags}) for a wide set of model parameters, and so the ratio $b_w / \det A$ is positive. This suggests that the behaviors shown in figures \ref{fig:pi_prevalences}.a) and \ref{fig:gamma_pi_diagrams} are robust and should remain for the whole coexistence region.

\subsection{Model B}

For model B, one can extract the dependence of the prevalences with $\pi$ (or $\chi$) directly from their analytical expressions (eqs. \ref{eq:solution_superinf_u} and \ref{eq:solution_superinf_v}), noticing that $\phi = 1 + \alpha \chi \lambda_2 / \lambda_1$ is an increasing function of $\chi$. The prevalence $v$ of disease II, as in equation \ref{eq:solution_superinf_v}, is clearly a decreasing function of $\phi$ for $\alpha > 1$ (which is the asymmetrically interacting case). From equation \ref{eq:solution_superinf_u}, we can also directly infer that the prevalence $u$ of disease I also decreases with $\phi$ for $s_2 > 0$ (or equivalently, $\lambda_2 \geq 1$). However, the coexistence phase (region (IV)) also comprehends a region at which $\lambda_2 < 1$, for which we should check the behavior with $\phi$ more carefully. Taking the partial derivative of $u$ with respect to $\phi$, we get:

\begin{equation}
    \label{eq:u_partial_phi}
    \frac{\partial u}{\partial \phi} = \frac{ -s_2 \left[ 1 + \phi (\alpha - 1) \right] - (s_1 - \phi s_2)(\alpha - 1) }{(1 + \phi(\alpha -1)^2)}
\end{equation}

From the above expression, the condition $\frac{\partial u}{\partial \phi} < 0$ can be simplified as:

\begin{equation}
    s_2 + (\alpha - 1) s_1 > 0,
\end{equation}
which is equivalent to the condition that $\lambda_2$ is above its critical value for coexistence, expressed by equation \ref{eq:superinf_lower_curve}. This means that $\frac{\partial u}{\partial \phi} < 0$ in the whole coexistence region and, therefore, both prevalences $v$ and $u$ are decreasing functions of $\phi$, $\chi$ and $\pi$. This is consistent with the observed behaviors in figures \ref{fig:pi_prevalences}.b) and \ref{fig:alpha_pi_diagrams}.


%

\end{document}